\documentclass[a4paper, 12pt]{article}
\usepackage{amsfonts, amsmath, amssymb, epsfig, subfigure, graphicx}

\usepackage[toc,page]{appendix}
\usepackage{authblk}
\usepackage{float}

\usepackage{tikz}
\usepackage{tabularx,ragged2e,booktabs,caption}
\usetikzlibrary{calc,patterns}
\usepackage{xcolor}
\usepackage{lineno}
\usepackage{multirow}
\newcommand{\commentout}[1]{}
 \footnotesep=\baselineskip
\def\wl{\par \vspace{\baselineskip}}

\usepackage{fancyhdr}
\begin{document}

%\linenumbers

\chead{}

\author{Valerio Capraro$^{1,*}$, Francesca Giardini$^{2,3*}$, Daniele Vilone$^4$, Mario Paolucci$^3$}
\affil{$^1$Department of Economics, Middlesex University Business School, NW44BT London, United Kingdom.\\ $^2$Department of Sociology, University of Groningen, Grote Rozenstraat 31
- 9712 TG Groningen, The Netherlands.\\ $^3$Laboratory of Agent Based Social Simulation (LABSS), Institute of Cognitive Sciences and Technology (ISTC-CNR), Rome, Italy.\\ $^4$Grupo Interdisciplinar de Sistemas Complejos (GISC), Departamento de Matem\'aticas, Universidad Carlos III de
Madrid, 28911 Legan\'es, Spain.\wl $^*$These authors contributed the same \wl Contact Author: V.Capraro@mdx.ac.uk}
\title{Partner selection supported by opaque reputation promotes cooperative behavior}
\maketitle

\begin{center}
\emph{Forthcoming in Judgment and Decision Making}
\end{center}

\pagebreak

\begin{abstract}
Reputation plays a major role in human societies, and it has been
proposed as an explanation for the evolution of cooperation. While the
majority of previous studies equates reputation with a transparent and
complete history of players' past decisions, in real life, reputations
are often ambiguous and opaque. Using web-based experiments, we
explore the extent to which opaque reputation works in isolating
defectors, with and without partner selection opportunities. Our
results show that low reputation works as a signal of
untrustworthiness, whereas medium or high reputation are not taken
into account by participants for orienting their choices. We also find
that reputation without partner selection does not promote cooperative
behavior; that is, defectors do not turn into cooperators only for the
sake of getting a positive reputation. Finally, in a third study, we
find that, when reputation is pivotal to selection, then a substantial
proportion of would-be-defectors turn into cooperators. Taken
together, these results provide insights on the characteristics of
reputation and on the way in which humans make use of it when
selecting partners but also when knowing that they will be selected.
\end{abstract}

\emph{keywords:} reputation, partner selection, cooperation, prisoner's dilemma, online transactions.

\pagebreak
\section*{Introduction} 

Partner selection, that is the ability to spot and preferentially interact with better social partners, is proposed to be a major factor in maintaining costly cooperation between individuals (No\"e and Hammerstein, 1994). Theories on the evolution of cooperation via indirect reciprocity (Nowak and Sigmund, 1998; Panchanathan and Boyd, 2004) emphasize the role of reputation in discriminating cheaters and supporting cooperation, and experimental studies using dynamic networks indeed suggest that human subjects tend to break links with defectors and form new links with cooperators (Rand, Arbesman and Christakis, 2011; Wang, Suri and Watts, 2012). 

Reputation-based partner selection requires the ability to evaluate others and to take into account third parties' evaluations, i.e., the result of being evaluated by others. A sensitivity towards others' presence and related evaluations is suggested by several studies. When subtle cues of being watched are present, like for instance a picture of watching eyes (Bateson, Nettle and Roberts, 2005), or even three dots in a face-like configuration (Rigdon et al., 2009), the probability of donating something significantly increases, both in laboratory experiments (Burnham and Hare, 2007; Haley and Fessler, 2005) and in field studies (Ernest-Jones, Nettle and Bateson, 2011; Yoeli et al., 2013). In economic games, cooperation increases when participants are informed about others' actions in a transparent and reliable manner, like for instance when they receive information about the amount of other players' past contributions in a Public Goods Game (Sommerfeld et al., 2007; Sommerfeld, Krambeck and Milinski, 2008), or in a two-players donation game (d'Adda, Capraro and Tavoni, 2015; Wedekind and Milinski, 2000). 

Although interesting, these studies present what Granovetter (1985) calls an `undersocialized' notion that equates reputation with a transparent and complete history of players' past decisions. In real life, reputations are based on personal evaluations, and they are often ambiguous, opaque and ephemeral. Nonetheless, humans strive to acquire positive reputations, but they also select partners and make decisions on the basis of partners' reputations. The
fragility of reputation is even more evident if we take into account
digital reputation, a widely used tool in online transactions and services.
According to Randy Farmer (2011, p. 16): ``Every top website is using
reputation to improve its products and services, even if only internally to
mitigate abuse. In short, reputation systems create real-world value''.
Reputation systems are designed to mediate and facilitate the process of
assessing reputations within specific communities (Dellarocas, 2011), and
they are built upon users' evaluations. (Dellarocas, 2012).

This kind of systems is pivotal to the establishment of online transactions among distant strangers characterized by asymmetric information, in which buyers have little or no information about the goods they are going to buy, as in electronic markets like eBay. In these systems, comments or feedbacks provided by previous users are essential to promote trust among parties, to overcome information asymmetries, and to minimize frauds (Diekman et al., 2014). However, in online reputation, evaluations are largely opaque in many ways. Sources are unknown, as well as their metrics, meaning that what someone rates as good can be below average for someone else, and this is especially true for ratings, like stars. Comments can be misleading too, or, even worse, fraudulent, because interested targets or their competitors can artificially manipulate reviews, and thus portray a very different situation (Matzat and Snjders, 2012). In spite of that, individuals heavily rely on others' reputations and evaluations, even when these are ambiguous and non transparent, as stars in online reputation systems (like Tripadvisor). 
Chevalier and Mayzlin (2006) examine book reviews on two online booksellers
and investigate the effects of reviews on books sales. Their analysis shows
that evaluations, both in the form of written reviews and average star
rankings, have an effect on book sales, even if negative and positive
reviews have asymmetric effects on consumers' behavior. Reputation systems are used in a variety of different contexts, from
philantropy to science (Masum, Tovey, 2011), but they all are based on
ambiguous, anonymous and usually aggregated evaluations. 

The aim of this work is to explore the extent
to which opaque reputations in the form of stars might support cooperation
in a strategic game. We consider both sides of evaluations, introducing
reputation-based partner selection in two variants: weak and strong. Given
our interest in understanding  ambiguous reputations, pervasive in online
environments, we choose to run  web-based experiments using the online
labor market Amazon Mechanical Turk (AMT). To avoid confounding factors related to
online transactions (prices, goods, sellers' features), to measure individuals' cooperative attitudes we decide to use a
standard one-shot Prisoner's Dilemma (Nowak, 2006; Perc \& Szolnoki, 2010; Capraro, 2013; Rand \& Nowak, 2013).
%We use a one-shot Prisoner's Dilemma because we consider it as more representative of the kind of interactions happening in the online world, which are usually risky and asynchronous. 

In order to single out the effects of evaluations and partner selection on individuals' behaviors, we designed three different studies. In the first experiment, we implemented a between-subjects design in order to understand how ambiguous evaluations (in the form of a grade obtained in a previous non-specified test) are taken into account, and how participants use such an ambiguous evaluation system when assessing their partners' behaviors. In the second and in the third study, we investigated the effect of knowing that one will be evaluated on one's own behavior, by increasing the consequences of being evaluated, ranging from none to the possibility of being selected for another round of the game by a third-party knowing only the person's reputation (see Methods for more details). 

Our results provide evidence of the importance of reputation even when it is opaque. More specifically, we report four major results: (i) people cooperate much less with low-reputation partners than with medium or high reputation partners, even if they do not know how that reputation was acquired, (ii) when given the opportunity to select a partner knowing only his or her reputation, people tend to select partners with high reputation; (iii) individuals use their own behavior as a baseline for evaluation, disregarding absolute value of actions; (iv) knowing that they will be evaluated by their partner and that a third party will have the opportunity to select them as future partners has the effect to turn a substantial proportion of defectors into cooperators.  

In sum, even when opaque and uncertain, reputation affects decision making: bad reputation works as a signal of anti-social behavior and, in combination with partner selection, promotes cooperative choices in digital
environments.
\section*{Methods}

We have conducted a series of three studies recruiting subjects through the online labour market Amazon Mechanical Turk (Mason and Suri, 2014; Paolacci and Chandler, 2014; Rand, 2012). We refer the reader to Table 1 for a summary of the three studies. Here we report the experimental design of each of the three studies. Full instructions are reported in the Appendix.

\begin{center}
\begin{tabular}{ |c|c|c|c|c|c|c|c|  }
   \hline   
   &\multicolumn{7}{  c |}{conditions}\\
\hline
Study 1 &baseline&low&neutral&high&evaluate&evaluate&evaluated\\
N = 544 &(no stars)&reputation&reputation&reputation&cooperator&defector&\\
&&(1 star)&(3 stars)&(5 stars)&&&\\
\hline
Study 2 &\multicolumn{3}{  c |}{external partner selection}&&&&\\
N = 96 &\multicolumn{3}{  c |}{}&&&&\\
\hline
Study 3 &\multicolumn{3}{  c |}{Random+Evaluated}&\multicolumn{3}{  c |}{Choose+Evaluated}&\\
N = 227 &\multicolumn{3}{  c |}{(Stage 1: random partner;}&\multicolumn{3}{  c |}{(Stage 1: choose partner;}&\\
 &\multicolumn{3}{  c |}{Stage 2: partner selection)}&\multicolumn{3}{  c |}{Stage 2: partner selection)}&\\
\hline
\end{tabular}\captionof{table}{Summary of our three experiments. We refer to the Methods section for more details and to the Appendix for full experimental instructions.}
 \end{center}

\wl
\textbf{\emph{Study 1}}

In this study we were interested in exploring whether opaque reputation is taken into account when interacting with someone, and how people apply an ambiguous reputation system when assessing their partner's behavior. After entering their TurkID, participants were randomly assigned to one of seven conditions. In the \emph{baseline} condition, participants were randomly matched to play a standard Prisoner's dilemma (PD). Specifically, each participant was given \$0.10 and had to decide whether to keep it (i.e., defect) or give it to the other player (i.e., cooperate). In the latter case, the \$0.10 would be multiplied by 2 and earned by the other player. After reading the instructions, participants were asked four general comprehension questions in random order. Participants failing any of the comprehension questions were automatically excluded from the survey.  Those who answered all comprehension questions were directed to the `decision screen', in which they could select either `keep' or `give', by means of appropriate buttons. %After making their decision, participants entered the demographic questionnaire (in which we asked for their age, gender, and level of education in a completely anonymous way), at the end of which they received the `survey code' needed to claim for their payment. 

Participants in the \emph{low reputation} condition played the same PD game as in the baseline condition, but, before making their choice, they were told that the person they were matched with had participated in a previous test (without receiving any information about the kind of test), in which he or she was rated 1 out of 5 stars. This information was initially presented in the `instructions screen' and then made salient in the `decision screen'. In reality, there was no previous test and, to compute the payoffs, we paired subjects at random. The \emph{neutral evaluation} condition was similar to the low reputation condition, with the only difference that participants were told that they were matched with a person who was rated 3 out of 5 stars in a previous test. In the \emph{high reputation} condition participants were told that they were matched with partners who was rated 5 out of 5 stars in a previous test.

Participants in the \emph{evaluated} condition played the same PD as in the baseline condition, but, before making their choice, they were informed that their choice would be communicated to the other participant, who would be asked to rate it from 1 to 5 stars. This procedure was real, as cooperators were paired with participants in the  \emph{evaluate cooperator} condition below, and defectors were paired with participants in the \emph{evaluate defector} condition below.

In the \emph{evaluate cooperator} condition participants first played the PD and then were informed that their partner had cooperated. At this stage, players were given the opportunity to rate the other participant's action from 1 to 5 stars. Finally, the \emph{evaluate defector} condition was similar to the \emph{evaluate cooperator} condition, with the only difference that participants were required to rate the behavior of an opponent who defected. 
\wl
\textbf{\emph{Study 2}}

As it will be shown in the Results section, the average cooperation in the \emph{evaluated} condition of Study 1 is statistically the same as the average cooperation in the \emph{baseline} condition, suggesting that the opportunity of being evaluated does not affect participants' choices in one-shot PD games, when the resulting reputation has no real consequences. We designed Study 2 in order to understand why
the reputation threat had had no effect on individuals' behaviors in Study
1. We introduced a light manipulation to the setting used in the previous
study, in which we informed players that their decisions (to cooperate or
to defect) would be communicated to the other player who could give a
rating going from 1 to 5 stars. Here, participants were also told that
ratings were collected with the purpose of creating a rank of Turkers among
which to select players with the highest ranks for further participation in
a particularly rewarding task. To build a reputation we used the data collected in the \emph{evaluate cooperator} and \emph{evaluate defector} conditions of Study 1. Specifically, each cooperator in this condition was randomly assigned to a participant in the \emph{evaluate cooperator} condition and was assigned the evaluation given by this particular participant. Similarly with defectors.

\wl
\textbf{\emph{Study 3}}

As it will be shown in the Results section, the light increase in the consequences of the evaluation experimented in Study 2 does not lead to an increase in cooperative behavior. The aim of Study 3 is to test whether a stronger form of partner selection would increase cooperative choices. To this end, we employed a two-stage game in order to test for differences in cooperation levels between the first and the second game. After entering their TurkID, participants were randomly assigned to either of two conditions. In the Random+Evaluated condition, the first stage consisted of a standard PD (as in the baseline condition in Study 1) played with a randomly selected partner. The following stage, instead, was divided in three parts: a game part, an evaluation part, and a selection part. In the game part, participants played another PD, neutrally framed, with a randomly selected person, denoted Person A, but they were told that the experimenter, in the next part of the stage (i.e., the evaluation part), will communicate their decision to another person, Person B (different from Person A), who was in charge of assigning participant's behavior a grade ranging from 1 to 5 stars.  Participants were also told that, in the third part (i.e., the selection part), another player (Person C) was given a list of 5 participants (including themselves), each characterized by a different grade, from which they could choose their partner for playing a PD. Participants were told that, in case they were selected by Person C, they would be playing another round of the PD with Person C. In reality, there was no other round. So, the total payoff of a participant was given by the sum of the payoffs obtained in the two PDs. Complete information about the three parts of Stage 2 was given all together at the beginning of the stage itself, and two comprehension questions (in addition to the four comprehension questions asked in Stage 1) were asked before participants were allowed to make their decision. 

The other treatment, Choose+Evaluated, was again in two stages. In the first stage, after reading the instructions of the PD, participants were told that they were grouped with other five participants, each of whom was characterized by a different number of stars obtained in a previous unspecified test. Participants were asked to select the participant with whom they wanted to play. In reality, this selection procedure was fictitious, and participants played with a randomly selected subject playing in the same condition, regardless of their selection. After the choices were made, participants enter the second stage, which was exactly the same as in the Choose+Evaluated condition.

\section*{Results}

A total of 962 subjects, located in the US, passed the comprehension questions and participated in our three studies. This corresponds to about 59\% of the total number of participants: about 41\% of subjects failed the attention check, and were automatically excluded from the survey. This is in line with previous studies using similar strategic situations. For instance, Capraro, Jordan and Rand (2014) report 32\% of subjects failing a very similar attention test. To avoid multiple observations from the same participant, each time we found a participant identified with the same IP address and/or the same TurkID, we kept only the first decision and eliminated the rest. As a consequence of this, the 962 participants that we analyze are \emph{distinct} in all measurable variables. Participants failing the comprehension questions in one study were allowed to participate in the subsequent studies.

\wl
\textbf{\emph{The effect of partner's opaque reputation on cooperative behavior}}

%\begin{figure}[H] %  figure placement: here, top, bottom, or page
   %\centering
   %\includegraphics[scale=0.70]{figure1.png} 
   %\caption{\emph{Representation of the six conditions of our Study 1: the baseline Prisoner's Dilemma (PD); the three PDs in which participants were given information about their partner's rating; and the two evaluate conditions, in which participants were asked to evaluate defectors and cooperators.}}
   %\label{fig:figure1}
%\end{figure}

We begin by analyzing how information about the other person's (opaque) reputation is taken into account when interacting with them. To this end, we analyze the data of the baseline (N = 96), the low reputation (N = 91), the neutral reputation (N = 87), and the high reputation (N = 82) conditions of Study 1. Results, summarized in Figure 1, show that participants cooperated much less with partners with low reputation (one star out of five) than with the others (Rank sum test. Low reputation vs baseline: $Z = -3.41$, $p = 0.0006$, effect size = 29\%; low reputation vs neutral reputation: $Z = -3.30$, $p = 0.0010$, effect size = 29\%; low reputation vs high reputation: $Z = -3.22$, $p = 0.0013$, effect size = 28\%;), even if they had no information about the way in which this reputation was acquired. However, there is no statistically significant difference between the rate of cooperation in the baseline condition and that in the `neutral' (Rank sum test: $Z = 0$, $p = 1$) and `high' (Rank sum test: $Z = 0.05$, $p = 0.9601$) reputation conditions. Thus, low reputation is a signal of anti-sociality, but high reputation is not a signal of pro-sociality. 

\begin{figure}[H] %  figure placement: here, top, bottom, or page
   \centering
   \includegraphics[scale=0.90]{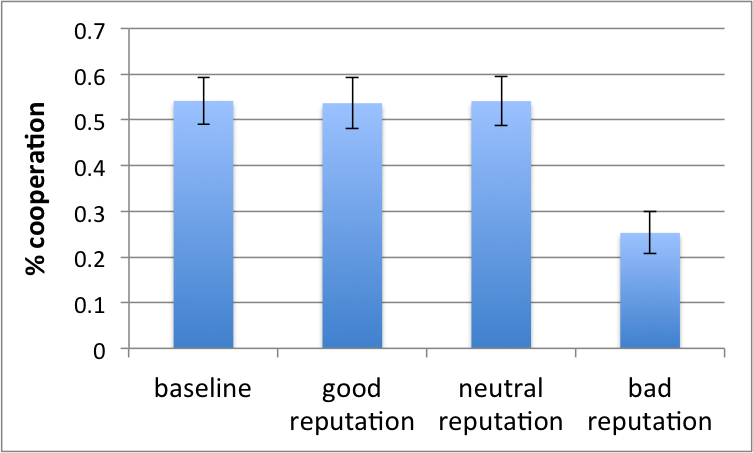} 
   \caption{\emph{Comparison among levels of cooperation when there is no information about the partner (baseline), or when they are ranked as having a high reputation (5 stars out of 5), a neutral one (3 stars) or a low reputation (1 stars). Error bars represent the standard error of the mean. Only low reputation seems to be informative for participants, who do not cooperate with people with low reputation, while no differences in cooperation are found between the three other evaluations.}}
   \label{fig:figure2}
\end{figure}

\wl
\textbf{\emph{The use of opaque reputation to assess others' behavior}}

Next, we analyze how subjects use opaque reputation to assess their partner's behavior. To this end, we analyze the data of the \emph{evaluate cooperation} (N = 95) and the \emph{evaluate defector} (N = 93) conditions of Study 1. Results, summarized in Figure 2, show that, when asked to assign a rate ranging from 1 to 5 stars to their partner, participants rated cooperators overwhelmingly higher than defectors. Specifically, the average grade of a cooperator was 4.91, while the average rate of a defector was 2.14 (Rank sum test: $Z=10.76$, $p < .0001$). Both defectors and cooperators gave cooperators very high rates. Indeed, the average grade of a cooperator when rated by another cooperator was 4.94, while the average grade of a cooperator when rated by a defector was 4.86 (Rank sum test: $Z = 0.64$, $p=0.5222$). On the other hand, cooperators evaluated defectors significantly worse than other defectors did: the average grade of a defector when rated by another defector was 2.80, while the average grade of a defector when rated by a cooperator was 1.48 (Rank sum test: $Z = 4.72$, $p < .0001$). The figure shows that some defectors were rated 4 or 5 stars. Data show that, in these cases, the evaluator was herself a defector. In other words, the maximum grade given to a defector by a cooperator was 3 stars. This means that evaluations were conditional on one's own behavior, and not based on the absolute positive or negative value of participants' choices. 

\begin{figure}[H] %  figure placement: here, top, bottom, or page
   \centering
   \includegraphics[scale=0.90]{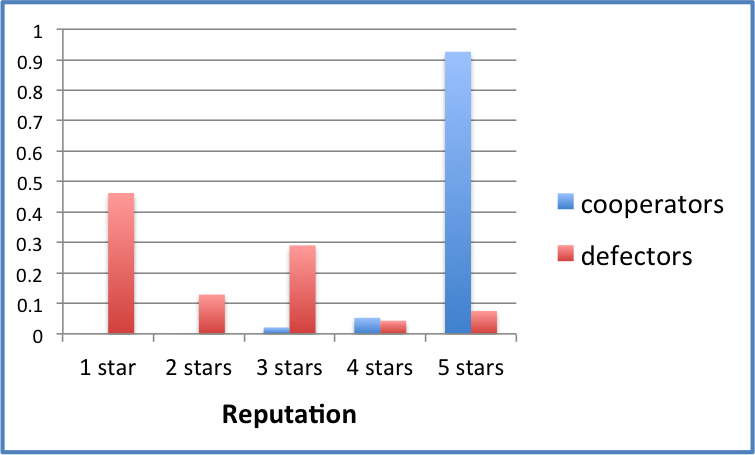} 
   \caption{\emph{Distributions of evaluations of cooperators and defectors on a scale from 1 to 5 stars. Cooperators received very positive evaluations, in contrast with defectors' grades.}}
   \label{fig:figure2}
\end{figure}

\wl
\textbf{\emph{The effect of being evaluated and external partner selection on cooperation}}

%\begin{figure}[H] %  figure placement: here, top, bottom, or page
  % \centering
   %\includegraphics[scale=0.70]{figure4.png} 
   %\caption{\emph{Graphical representation of the two conditions of Study 2. They differ in the way the rating obtained by player A will be used.}}
   %\label{fig:figure3}
%\end{figure}

Next, we analyze the effect of being evaluated on cooperative behavior in two cases: when the evaluation phase is not followed by real partner selection; and when it is followed by \emph{external partner selection}, that is, by the possibility of being selected by the experimenter for new studies. To this end, we analyze the data of the \emph{evaluated} condition of Study 1 (N = 95) and Study 2 (N = 96). We compare the results of Study 2 with those of the \emph{evaluated} and the \emph{baseline} conditions in Study 1, although these experiments were conducted in different times, only to understand whether the light increase in the consequences of the evaluation in Study 2 is likely to produce relevant changes in cooperative behavior. Figure 3 summarizes the results and shows that neither treatments had a significant effect on cooperative behavior (Rank sum test. Baseline vs Evaluated: $Z = -0.32$, $p = 0.749$; baseline vs study 2: $Z = -0.57$, $p = 0.453$). 

\begin{figure}[H] %  figure placement: here, top, bottom, or page
   \centering
   \includegraphics[scale=0.60]{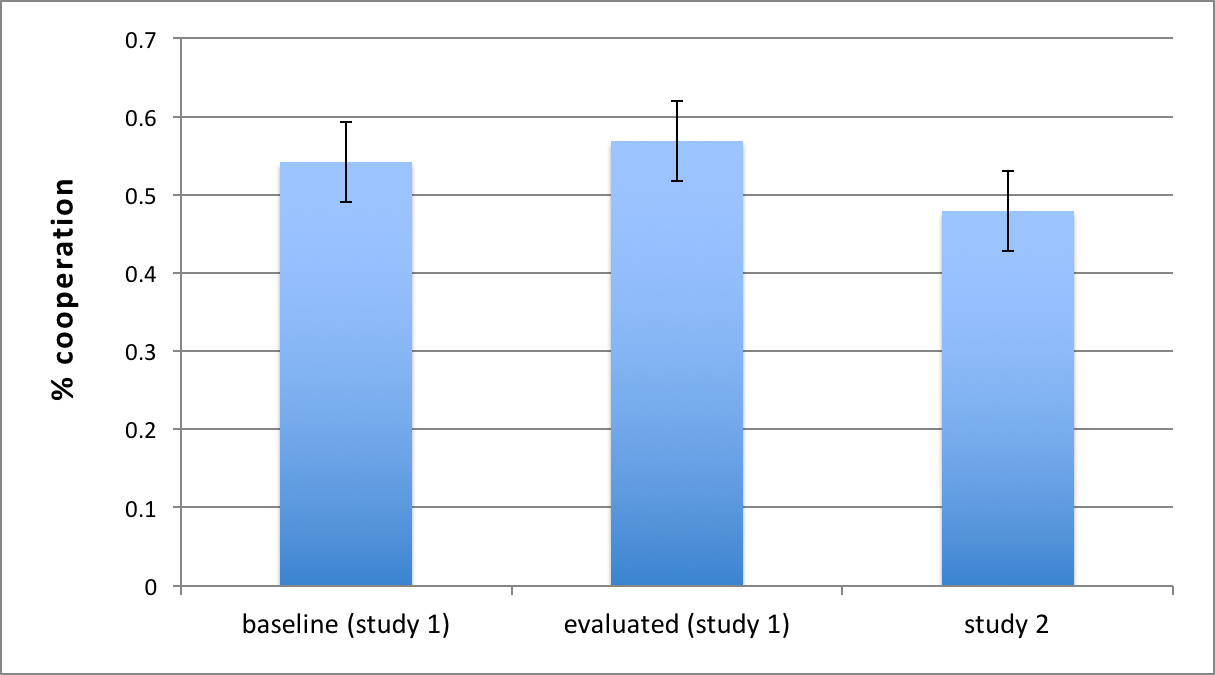} 
   \caption{\emph{Percentage of cooperation for the evaluation without partner selection condition (evaluated, Study 1) and for the evaluation with possibility to be selected by the experimenter for another study (Study 2). Error bars represent the standard error of the mean.}}
   \label{fig:figure5}
\end{figure}

\wl
\textbf{\emph{The effect of being evaluated and internal partner selection on cooperation}}

%\begin{figure}[H] %  figure placement: here, top, bottom, or page
  % \centering
   %\includegraphics[scale=0.75]{figure6.png} 
   %\caption{\emph{Graphical representation of the two conditions used in Study 3.}}
   %\label{fig:figure6}
%\end{figure}

Finally, we analyze the effect of being evaluated on cooperative behavior, when the evaluation is followed by \emph{internal partner selection}, that is, the possibility of being selected by another participant for playing another round of the PD. To this end, we analyze the data of Study 3 (Random+Evaluated: N = 104; Choose+Evaluated: N = 123). Figures 4 and 5 show an increase of cooperation in the second PD game with respect to the first one, in both experimental conditions. Cooperation increased both when partners were randomly assigned (Random+Evaluated), and when participants chose with whom to interact (Choose+Evaluated), suggesting that the opportunity of being selected as a partner in the next game increased cooperative choices. To understand what drives this increase in cooperative behavior, we do a within-subject analysis looking at those subjects who changed strategy from the first PD to the second PD. In both experimental conditions, we find that virtually all of those subjects who cooperated in Stage 1 remained cooperators in Stage 2, whereas a substantial proportion of subjects who defected in Stage 1 became cooperative in Stage 2. Specifically, in the \emph{random+evaluated} condition, we find that 55 participants cooperated and 49 defected in the first PD. Among these cooperators, only 3 of them changed strategy and defected in the second PD. On the other hand, among the defectors, 40\% of them (20 out of 49) changed strategy and cooperated in the second PD. Similarly, in the \emph{choose+evaluated} condition, 69 participants cooperated and 53 defected in the first PD. Among these cooperators, none of them changed strategy, that is, all of them cooperated also in the next PD. Among the defectors, about 30\% (16 out of 53) changed strategy and cooperated in the second PD. Thus, in both cases, we found that the combination of reputation and partner choice was effective in turning a substantial proportion of defectors into cooperators. %Multilinear regression shows that there is no effect of socio-demographic variables on the probability of changing strategies (all p?s > 0.2).

\begin{figure}[H] %  figure placement: here, top, bottom, or page
   \centering
   \includegraphics[scale=0.90]{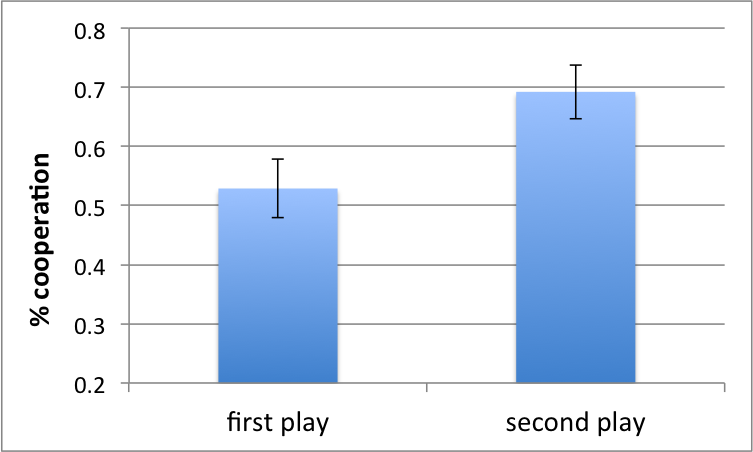} 
   \caption{\emph{Percentage of cooperative choices in the first and second stage of the Random+Evaluated condition. Error bars represent the standard error of the mean. In the second stage, participants played the PD knowing that their choice would be evaluated by another person and that a third party could select them, for playing another round of the PD, from a list of five participants, one for each possible grade. This significantly increased cooperative choices, compared to the first, completely neutral, PD.}}
   \label{fig:figure4}
\end{figure}

\begin{figure}[H] %  figure placement: here, top, bottom, or page
   \centering
   \includegraphics[scale=0.90]{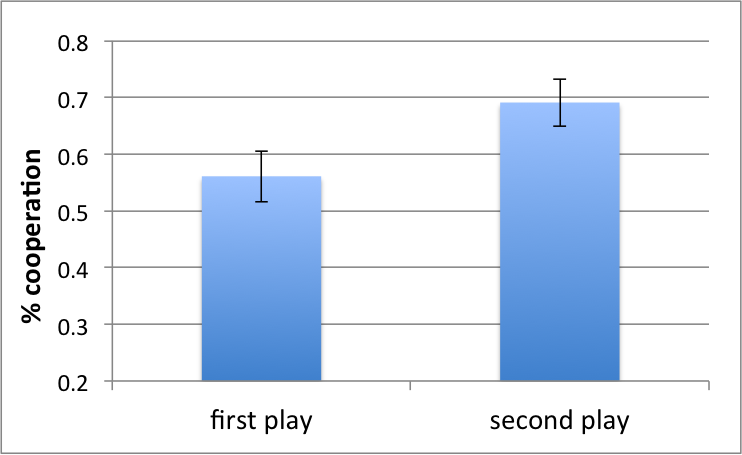} 
   \caption{\emph{Percentage of cooperative choices in the first and second stage of the Choose+Evaluated condition. Error bars represent the standard error of the mean. Even when choosing a partner, participants did not show increases in cooperation until the second game, when they were evaluated and, possibly, selected for another game.}}
   \label{fig:figure5}
\end{figure}

Finally, we investigated people's preferences when they had the opportunity to choose a partner for playing the first PD (i.e., in the \emph{choose+evaluated} condition). We find that the majority of people, but not all (75\%), preferred to play with a partner with 5 stars. Those who decided to play with an opponent with less than 5 stars were significantly less cooperative than those who decided to play with an opponent with 5 stars (average cooperation: 35\% vs 63\%, Rank sum, $p=0.022$). 

\section*{Discussion}

In this experimental work, we focused on the role of evaluations on cooperative behaviors, disentangling the effects of evaluating others from those of being evaluated, and assessing the relative importance of partner selection in these processes. The role of reputation on cooperative behaviors in social dilemmas is well-established (Alexander, 1987; Nowak and Sigmund, 2005; Rand and Nowak, 2013; Nax, Perc, Szolnoki and Helbing, 2015), but less is known on the consequences of reputation format and presence or absence of partner selection, especially in the noisy and ambiguous online environment. 

We tested whether a reputation, low, neutral or high, coming from a completely unknown source and acquired in an obscure situation (i.e., what we termed opaque reputation) was used in a web-based experiment by participants who had to decide whether to cooperate or defect in a one-shot Prisoner Dilemma. Cooperation rates are affected by partners'
rankings (expressed as one, three or five stars out of five), but this
effect is not symmetrical. Participants rated with one star received
significantly less cooperation than participants with neutral (three
stars) and high (five stars) reputations. 

The preminence of bad
evaluations over good ones is a distinguishing trait of human psychology
(Baumeister et al., 2001), and it seems especially salient in social
contexts. Anderson and colleagues (2011) show that negative gossip
associated with neutral faces dominates longer in a visual discrimination
task. In the online world, Chevalier and Mayzlin (2006) find that a
negative review of a book has a stronger influence than a positive one, in
online book sellers websites like Amazon. Our results are thus in line with previous findings.

In the second study, participants were told that their actions would have been evaluated using the same opaque reputation system used in the first study, but our results showed that evaluation alone, without any actual partner selection, was not effective in increasing cooperative choices. Informing participants that their partners had the opportunity to rate them did not make cooperation increase with respect to a baseline, not even when we told them that a general ranking of players with the possibility of participating in future rewarding tasks would be created. This result allows us to narrow down the effect of partner selection, while stressing the fact that cooperation is not enhanced by the simple awareness that someone evaluates us. 

In order to understand better how and under which conditions partner selection is effective in promoting cooperation, we design a third study using a repeated game in which participants were told that partner selection was real. Our results show an increase in cooperation levels between the first game (without evaluation and partner selection) and the second one, in which players were told that their behavior would have been evaluated and this evaluation transmitted to another potential player. The increase in cooperation was not due to `active partner choice', because it happened also when partners were randomly assigned, therefore leading us to conclude that the combination of being evaluated and being selected explained the observed increase in cooperation. In the third experiment, we also observed that a subset of participants selected partners with less than five stars and did not cooperate with them. 

Although centered on digital reputation, our
work adds to the literature on reputation systems but also to the general
literature on cooperation. Studies on web-based reputation systems usually
analyze real websites designed for enhancing trust in online transactions,
in which real goods or services are exchanged (Dellarocas, 2006; Farmer and
Glass, 2010). Our studies use a completely neutral setting, in which
motivations related to individual preferences or needs do not enter
decision making, but the main elements of reputation systems, i.e.,
evaluations and partner selection, are present. Showing that an opaque
negative reputation can support cooperation complements evolutionary
accounts that consider negative reputation as a powerful means for
detecting and avoiding cheaters (Barkow, 1992; Giardini and Conte, 2012;
Hess and Hagen, 2006; Nowak and Sigmund, 1998), a view supported also by
several experimental findings (Anderson et al., 2011; Feinberg et al, 2012;
Sommerfeld, Krambeck and Milinski, 2008). Reputation alone, however, seems
to have no effect when it is devoid of partner choice, showing that what is
called the ``threat of gossip'' (Beersma and Van Kleef, 2011; Piazza and
Bering, 2008) is effective only when consequences of reputation are evident.

Reputation plays a major role in human sociality, and it has been proposed as an explanation for the evolution of costly cooperation. In recent years, reputation has become central also in online systems, even if it is much less controllable and completely opaque. Our findings suggest that reputation, even if opaque, works in isolating defectors, but its value is conditional on participants' behaviors. Moreover, when partner selection is not effective, individuals do not become more cooperative only for the sake of getting a positive reputation, at least not in an anonymous online environment. On the contrary, when reputation is pivotal to selection, it leads individuals to change their behaviors and to cooperate. The behavioral switch is strong: the mere possibility of being selected for a new interaction turn about 35\% of defectors into cooperators. This finding is interesting since it suggests yet another way to promote cooperative behavior in the field (see Kraft-Todd et al. 2015, for a recent review on interventions to promote cooperation in the field). 

More generally, our experiments provide insights on the way in which humans use reputational information in uncertain environments, like for instance in online interactions. This has implications that exceed online markets and can be applied to several domains. For example, companies or universities, whose success is highly based on cooperation among their employees, might develop a reputational system, according to which colleagues that have been working together on the same project can rate one another.

Our experiments certainly have some limitations. Study 1 uses deception in three out of seven conditions; specifically, those in which participants are told that their partners obtained a certain grade in a previous unspecified test. In reality there was no previous test. Study 3 uses deception when subjects are told that they could be selected for another round of the PD, depending on how their choice would be evaluated by a third party. In reality, although the evaluation procedure was real, there was no selection for other rounds. In general, the use of deceptive messages leads to a decrease of the effect size (when there is a true effect), driven by a proportion of subjects that  may anticipate the fact that the manipulation is not real. Thus, the effect sizes that we found in Study 1 and Study 3 are likely to be a lower bound for the true effect sizes. Understanding the true sizes of these effects is an important direction for future work.
On the other hand, it is possible that the asymmetric effect of reputation information on cooperative behavior (Study 1) is an artifact of the use of deceptive messages: participants paired with high reputation partners may be more skeptical about the reality of the manipulation than those paired with low reputation partners. Although, as discussed above, this asymmetry is in line with previous studies, we cannot exclude that, in our case, it is driven by the use of deception and thus we leave this question for further research.

\section*{Acknowledgment}

D.V. received support from H2020 FETPROACT-GSS CIMPLEX Grant No. 641191.

\pagebreak

\pagebreak
\begin{appendix}

\section{Experimental instructions}

\par Each study started with the same two screens. In the first screen we asked participants to type their WorkID, while in the second screen we informed them about the average length of the study, the corresponding participation fee, and the fact that there would be comprehension questions. They were also informed that they would be automatically excluded from the survey in case they fail any of them. At the end of this screen, participants could either continue and play, or end the survey. 
After each treatment, standard demographic questions were asked, at the end of which participants were given the completion code needed to claim for the payment.
We report below full instructions of each study.
\wl
\textbf{\emph{Study 1}}

In Study 1, participants were randomly assigned to play one of six conditions (Baseline, Low reputation, Neutral reputation, High reputation, Evaluate cooperator, and Evaluate defector). Instructions of these conditions were as follows (we report the comprehension questions only in the Baseline condition, but they were present in each condition).
\wl
\emph{Baseline}

You have been paired with another, anonymous participant. How much money you earn depends on your own choice, and on the choice of the other participant.
 
You are given 10 additional cents. You can either keep the money or give it to the other participant. If you decide to give the money, your 10c will be multiplied by 2 and earned by the other participant.
 
The other participant will be given the same choice.

So:
\begin{itemize}
\item if you both give, you both get 20 cents
\item if you both keep, you both get 10 cents 
\item if you give and the other person keeps, then you earn nothing
\item if you keep and the other participant gives, then you get 30 cents.
\end{itemize} 
The other person is REAL and will really make a decision. Once you have each made your decision, neither of you will ever be able to affect each others' bonuses in later parts of the HIT.

Now we will ask you several questions to make sure that you understand how the payoffs are determined.

YOU MUST ANSWER ALL THESE QUESTIONS CORRECTLY TO RECEIVE A BONUS!

Which action by YOU gives YOU a higher bonus?
\begin{itemize}
\item Keep
\item Give
\end{itemize}
Which action by YOU gives the OTHER PLAYER a higher bonus?
\begin{itemize}
\item Keep
\item Give
\end{itemize}

Which action by the OTHER PLAYER gives the OTHER PLAYER a higher bonus?
\begin{itemize}
\item Keep
\item Give
\end{itemize}

Which action by the OTHER PLAYER gives YOU a higher bonus?
\begin{itemize}
\item Keep
\item Give
\end{itemize}

Congratulations! You passed all comprehension questions. It is now time to make a decision.

WHAT IS YOUR CHOICE?
\begin{itemize}
\item Keep
\item Give
\end{itemize}
\wl
\emph{Low reputation}

You have been paired with another participant. 
 
The other participant is not completely anonymous. In a previous study, he or she participated in a test. We will tell you how he or she was rated in this test later.
 
You are now paired with this person and you both have to make a choice. How much money you earn depends on your own choice, and on the choice of the other participant.  
 
You are given 10 additional cents. You can either keep the money or give it to the other participant. If you decide to give the money, your 10c will be multiplied by 2 and earned by the other participant.
 
The other participant will be given the same choice.

So:
\begin{itemize}
\item if you both give, you both get 20 cents
\item if you both keep, you both get 10 cents 
\item if you give and the other person keeps, then you earn nothing
\item if you keep and the other participant gives, then you get 30 cents.
\end{itemize}
The other person is REAL and will really make a decision. Once you have each made your decision, neither of you will ever be able to affect each others' bonuses in later parts of the HIT.

(Comprehension questions were asked here)

Congratulations! You passed all comprehension questions. It is now time to make a decision.

YOU HAVE BEEN PAIRED WITH A PARTICIPANT RATED 1 STAR OUT OF A MAXIMUM OF 5.

What is your choice?
\begin{itemize}
\item Keep
\item Give
\end{itemize}

\wl
\emph{Neutral reputation}

This condition was identical to the Low reputation condition, with an important difference: the sentence `YOU HAVE BEEN PAIRED WITH A PARTICIPANT RATED 3 STARS OUT OF A MAXIMUM OF 5' was replaced by this sentence `YOU HAVE BEEN PAIRED WITH A PARTICIPANT RATED 3 STARS OUT OF A MAXIMUM OF 5', in order to manipulate the partner's reputation. 
\wl
\emph{High reputation}

This condition was identical to the Low reputation condition, but the partner had a very positive reputation, as expressed in the following sentence: `YOU HAVE BEEN PAIRED WITH A PARTICIPANT RATED 5 STARS OUT OF A MAXIMUM OF 5'.

\wl
\emph{Evaluate cooperator}

You have been paired with another, anonymous participant. How much money you earn depends on your own choice, and on the choice of the other participant.
 
You are given 10 additional cents. You can either keep the money or give it to the other participant. If you decide to give the money, your 10c will be multiplied by 2 and earned by the other participant.
 
The other participant will be given the same choice.

So:
\begin{itemize}
\item if you both give, you both get 20 cents
\item if you both keep, you both get 10 cents 
\item if you give and the other person keeps, then you earn nothing
\item if you keep and the other participant gives, then you get 30 cents.
 \end{itemize}
The other person is REAL and will really make a decision. Once you have each made your decision, neither of you will ever be able to affect each others' bonuses in later parts of the HIT.

(Comprehension questions were asked here)

Congratulations! You passed all comprehension questions. It is now time to make a decision.

WHAT IS YOUR CHOICE?
\begin{itemize}
\item Keep
\item Give
\end{itemize}

The other participant decided to GIVE. Please rate his or her behavior.
\begin{itemize}
\item 1 star
\item 2 stars
\item 3 stars
\item 4 stars
\item 5 stars
\end{itemize}
\wl
\emph{Evaluate defector}

This condition was identical to the EvaluateC condition, except for the fact that the partner decided to keep, therefore the word `GIVE' in the last screen was replaced by `KEEP'.

\wl
\textbf{\emph{Study 2}}

In Study 2 participants were randomly selected to participate in either of two conditions: weak priming and strong priming. Below we report exact instructions of the treatment. Comprehension questions were exactly the same as in Study 1, so we do not report them again.
\wl
\emph{Weak priming}

You have been paired with another, anonymous participant. How much money you earn depends on your own choice, and on the choice of the other participant.
 
IMPORTANT: AFTER YOU MAKE YOUR CHOICE, WE WILL COMMUNICATE IT TO THE OTHER PARTICIPANT, WHO WILL BE GIVEN THE OPPORTUNITY TO RATE YOUR BEHAVIOUR GIVING 1 TO 5 STARS.

You are given 10 additional cents. You can either keep the money or give it to the other participant. If you decide to give the money, your 10c will be multiplied by 2 and earned by the other participant.
 
The other participant will be given the same choice.

So:
\begin{itemize}
\item if you both give, you both get 20 cents
\item if you both keep, you both get 10 cents 
\item if you give and the other person keeps, then you earn nothing
\item if you keep and the other participant gives, then you get 30 cents.
\end{itemize} 
The other person is REAL and will really make a decision. Once you have each made your decision, neither of you will ever be able to affect each others' bonuses in later parts of the HIT.

(Comprehension questions were asked here)

Congratulations! You passed all comprehension questions. It is now time to make a decision.

REMEMBER THAT, AFTER THE CHOICES ARE MADE, THE OTHER PARTICIPANT WILL BE GIVEN THE OPPORTUNITY TO RATE YOUR BEHAVIOUR.

What is your choice?
\begin{itemize}
\item Keep
\item Give
\end{itemize}

\emph{Strong priming}

You have been paired with another, anonymous participant. How much money you earn depends on your own choice, and on the choice of the other participant.
 
You are given 10 additional cents. You can either keep the money or give it to the other participant. If you decide to give the money, your 10c will be multiplied by 2 and earned by the other participant.
 
The other participant will be given the same choice.

So:
\begin{itemize}
\item if you both give, you both get 20 cents
\item if you both keep, you both get 10 cents 
\item if you give and the other person keeps, then you earn nothing
\item if you keep and the other participant gives, then you get 30 cents.
\end{itemize} 
The other person is REAL and will really make a decision. Once you have each made your decision, neither of you will ever be able to affect each others' bonuses in later parts of the HIT.

(Comprehension questions were asked here)

Congratulations! You passed all comprehension questions.

BEFORE YOU MAKE A DECISION, WE INFORM YOU THAT WE ARE DEFINING A RATING SYSTEM FOR PARTICIPANTS. YOUR BEHAVIOR WILL BE RATED BY OTHER PARTICIPANTS (FROM 1 TO 5 STARS) AND THIS INFORMATION WILL BE STORED IN OUR DATABASE AND USED FOR SELECTING PARTICIPANTS IN FURTHER TASKS.

What is you choice?
\begin{itemize}
\item Keep
\item Give
\end{itemize}

\wl
\textbf{\emph{Study 3}}

In Study 3, participants were randomly assigned to either of two conditions: Random+Evaluated and Choose+evaluated. Full instructions are reported below. Comprehension questions about the Prisoner's Dilemma were exactly the same as in the previous studies and so we do not report them again.
\wl
\emph{Random+Evaluated}

This HIT is divided in two stages.

In this first stage, you are given 10 additional cents. You can either keep it or give it to the other participant. If you decide to give it, your 10c will be multiplied by 2 and earned by the other participant.
 
The other participant will be given the same choice.

So:
\begin{itemize}
\item if you both give, you both get 20 cents
\item if you both keep, you both get 10 cents 
\item if you give and the other person keeps, then you earn nothing 
\item if you keep and the other participant gives, then you get 30 cents. 
\end{itemize}
The other participant is REAL and will really make a choice. This is a one-shot interaction. In the second stage of this HIT you will be grouped with other participants. The current participant will not have the possibility to influence your bonus in later parts of the HIT. 

(Comprehension questions were asked here)

Congratulations! You passed all comprehension questions.
 
It is now time to make a choice. What do you want to do?
\begin{itemize}
\item Keep
\item Give
\end{itemize}
The second stage of this HIT consists of three parts:

PART 1

Here you will play the same game as in the first stage, with a random participant. You know nothing about him or her. 
 
Recall, briefly, the rules of the game: You are given 10 additional cents. You can either keep it or give it to the other participant. If you decide to give it, your 10c will be multiplied by 2 and earned by the other participant. The other participant is given the same choice.\\
 
\par PART 2

Here we will communicate your choice to another participant, Person B (different from Person A). The role of Person B is to rate your choice by giving it a score ranging from 1 to 5 stars.\\
 
\par PART 3

Here we will show to another participant, Person C (different from Persons A and B) the number of stars you received from Person B. Person C will choose with whom to play from a list of 5 participants, including you, each one characterized by a score. If Person C chooses to play with you, you will play again and you will have the opportunity to win more money. Otherwise, if Person C chooses to play with someone else, your HIT will end.

Now we will ask you two simple comprehension questions in order to make sure you understood the procedure. Recall that you must answer these questions correctly in order to get a bonus.  \\

What happens in Part 2?
\begin{itemize}
\item I will play the same game again
\item The choice I made in Part 1 will be communicated to a third-party, who will rate it
\item The choice I made in Part 1 will be communicated to the same person with whom I played in Part 1, who will rate it
\item The choice I made in Part 1 will be communicated to the same person with whom I played in Part 1, after which he or she makes his or her choice.\\
\end{itemize}

What happens in Part 3?
\begin{itemize}
\item I will play the same game again
\item I will play the same game with a person who knows the choice I made in Part 1
\item I will play the same game with a person who knows how the choice I made in Part 1 was rated in Part 2
\item I will be grouped with a person who knows how my choice was rated and decides whether to play with me or not.
\end{itemize}
Congratulations! You passed all comprehension questions.

What is your choice?
\begin{itemize}
\item Keep
\item Give
\end{itemize}
\wl
\emph{Choose+evaluated}

This HIT is divided in two stages.

In this first stage, you are given 10 additional cents. You can either keep it or give it to the other participant. If you decide to give it, your 10c will be multiplied by 2 and earned by the other participant.
 
The other participant will be given the same choice.

So:
\begin{itemize}
\item if you both give, you both get 20 cents
\item if you both keep, you both get 10 cents 
\item if you give and the other person keeps, then you earn nothing 
\item if you keep and the other participant gives, then you get 30 cents. 
\end{itemize}
The other participant is REAL and will really make a choice. This is a one-shot interaction. In the second stage of this HIT you will be grouped with other participants. The current participant will not have the possibility to influence your bonus in later parts of the HIT. 

(Comprehension questions were asked here)

Congratulations! You passed all comprehension questions.
 
You have been grouped together with other five participants. In a previous HIT, these people participated in a test. They rated as follows:

\begin{itemize}
\item Person A's grade is 1 star out of a maximum of 5
\item Person B's grade is 2 stars out of a maximum of 5
\item Person C's grade is 3 stars out of a maximum of 5
\item Person D's grade is 4 stars out of a maximum of 5
\item Person E's grade is 5 stars out of a maximum of 5
\end{itemize}

Please choose the participant you would like to play with: 
\begin{itemize}
\item Person A (grade: 1 star out of a maximum of 5)
\item Person B (grade: 2 stars out of a maximum of 5)
\item Person C (grade: 3 stars out of a maximum of 5)
\item Person D (grade: 4 stars out of a maximum of 5)
\item Person E (grade: 5 stars out of a maximum of 5)
\end{itemize}
It is now time to make a choice. What do you want to do?
\begin{itemize}
\item Keep
\item Give
\end{itemize}
(The second stage of this condition was identical to the second stage of the `Random+Evaluated' condition)



\end{appendix}

\end{document}